\documentclass{PoS}

\usepackage{epsfig}
\usepackage{bm}
\usepackage{amssymb}
\usepackage{fontenc}
\usepackage{times}
\usepackage{mathptmx}
\usepackage{graphicx}

\title{Exploring the chiral singularities of two color lattice QCD at strong coupling}

\ShortTitle{Two color QCD}

\author{\speaker{Fu-Jiun Jiang}\\
        Box 90305, Duke University, Durham NC 27708, USA\\
        E-mail: \email{fjjiang@phy.duke.edu}}
\abstract{
We use a modified directed path algorithm to study the finite temperature chiral singularities of two color lattice QCD with
staggered fermions at strong coupling. Our lattice calculations are done at a fixed finite temperature in the broken phase 
with a variety of different quark masses. We find that to the lowest order, the behavior of our observables, namely the condensates and the decay constants are all consistent with the predictions of the low energy effective theory of our model. 
We further notice that in order to see consistency, the quark masses used in our lattice calculations need to be quite small 
than typically used.}

\FullConference{XXIVth International Symposium on Lattice Field Theory\\
                July 23-28, 2006\\
                Tucson, Arizona, USA}

\begin{document}

\section{INTRODUCTION}
To study the chiral singularities in lattice QCD is a challenge due to a variety of technical difficulties, for example, the lack of algorithms which work efficiently for 
the simulations with realistic pion mass and the logarithmic nature of the chiral singularities. Because of these difficulties, today most lattice simulations are performed using large quark masses and the results are then extrapolated to the physical regime by using chiral perturbation theory ($ChPT$). However, some attempts to connect lattice QCD with the predictions from $ChPT$ do not give clear answers \cite{chilog1:2003ab,chilog2:2003ab,chilog3:2003ab,chilog4:2003ab,chilog5:2003ab}. 
Given the difficulties described above in understanding the chiral singularities of realistic QCD simulations, we focus on a simpler toy model, namely the two color lattice QCD at strong couplings. Further, since it is easier to detect power-like singularities as compared to logarithmic ones, we have done our lattice calculations at a fixed finite temperature in the broken phase. In order to study the finite temperature singularities of the model, we have developed an efficient algorithm based on the directed path algorithm \cite{Adams:2003cc}. With this algorithm, the singularities of our model can be studied quantitatively and the results will be present in the following.. 

\section{The Model and the Symmetries}
\label{model}
The lattice action of two color lattice QCD (2CLQCD) model at zero baryon density which we have studied is given by
\begin{equation}
S=-\sum_{x,\alpha}r_\alpha \eta_{\alpha}(x)
\Bigg[ 
\overline{\chi}(x)U_{\alpha}(x)\chi(x+\hat{\alpha})
- \overline{\chi}(x+\hat{\alpha})U_{\alpha}^{\dagger}(x)\chi(x) + m\overline{\chi}\chi(x)\Bigg].
\label{scact}
\end{equation}
Here $m$ is the quark mass and the quark fields $\overline{\chi}(x)$ and $\chi(x)$ are two 
components vector fields with Grassmann variables. The gauge fields $U_{\alpha}(x)$ are elements of $SU(2)$ group and live on 
the links between $x$ and $x+\hat{\alpha}$ where $\alpha=t,1,2,3$. The
factor $r_\alpha=1$ for $\alpha=1,2,3$ and $r_t = 1/a_t$. By choosing an $L_t\times L^d$ 
lattice (periodic in all directions) we can study thermodynamics in
the $L \rightarrow \infty$ limit at a fixed $L_t$ by defining $T=1/(a_t)^2$ 
as the parameter that represents the temperature of the model. The absence of the gauge 
action shows that we are in the strong gauge coupling limit.
A detailed discussion of the symmetries of 2CLQCD can be found 
in \cite{Hands:1999md,Nishida:2003uj}. Here we only review them 
briefly. When $m = 0$ our model has a $U(2)$ global symmetry. In particular, this
$U(2)$ global symmetry contains a $U_{B}(1)$ baryon number symmetry and a $U_{C}(1)$ chrial symmetry. At low temperature, the $U(2)$ global symmetry is spontaneously broken down to $U_B(1)$ which leads to the existence of three massless Goldstone
particles. In the presence of non-zero quark masses, only the baryon number symmetry $U_{B}(1)$ survives and remains unbroken\footnote{This $U_{B}(1)$ baryon number symmetry will survive until the chemical potential $\mu$ is switched on and reachs a critical value $\mu_{C}$.}. Further, once massive 
quarks are present, the three Goldstone particles acquire masses. One we call ``pion'' 
and the other two, which are degenerate in mass are called ``diquark baryons''.

\section{Dimer Representation of the Model}
\label{dimer}
One of the computational advantages of the strong coupling limit is that
in this limit it is possible to rewrite the partition function as a sum over configurations containing gauge invariant objects, namely 
monomers, dimers and baryons \cite{Rossi:1984cv,Klatke:1989xy}. A simple calculation shows that the partition function $Z$ of the model can be rewritten as:
\begin{equation}
Z=\sum_{\{ K\}}\prod_x\sum_{i}\frac{m^{n_x}}{n_x!}(\frac{1}{2})^{[k_{i \neq t}(x) + \mid b_{i \neq t}(x) \mid]}T^{[k_{t}(x) + \mid b_{t}(x) \mid]}.
\label{dimer1}
\end{equation}
where the variables $n_{x}=0,1,2$, $k_{l}(x)=0,1,2$, $b_{l}(x) = 1,0,-1$, $l = t,1,2,3$ associated with each
lattice site $x$ are defined as the number of monomers, dimer bonds and oriented baryon bonds connecting lattice sites $x$ and $x+\hat{l}$ respectively. Further, the baryon bonds always form isolated self-avoiding closed loops and each lattice site $x$ in the remaining configuration without baryon loops satisfies the constraint: $\sum_l k_{l}(x)+n_{x} = 2$. Note that the partition function in (\ref{dimer1}) is positive definite, hence a Monte-Carlo algorithm
can in principle be designed to study this system. The algorithms used in our lattice simulations is an extension of the directed path
algorithm discovered in \cite{Adams:2003cc}. We refer readers to \cite{FJ:2006ab} and do not provide the details of this algorithm here.

\section{Observables}
A variety of observables can be measured with our algorithm. We will focus
on the following: (1) The chiral two point function and the chiral susceptibility $\chi_{C}$
which are given by $G_C(z,z') = \Bigg \langle \bar{\chi}(z)\chi(z) \bar{\chi}(z')\chi(z')
\Bigg \rangle$ and $\chi_{C}\equiv \sum_{z'} G_C(z,z')$.
 (2) The diquark two point function which is given by $G_B(z,z') = \Bigg \langle 
\chi_1(z)\chi_2(z) \bar{\chi}_2(z')\bar{\chi}_1(z') \Bigg \rangle$. (3) The helicity modulus associated with the $U(1)$ 
chiral symmetry $ Y_{C} \equiv
\frac{1}{3V_s}\sum_{\mu=1,2,3}
\Bigg \langle (\sum_{x}A_{\mu}(x))^{2} \Bigg \rangle
\label{yc}
$ where $A_{\mu}(x)= \varepsilon(x) \left[ |b_\mu(x)| + k_\mu(x)\right]$
and $V_s = L^3$. (4) The helicity modulus associated with the $U(1)$ 
baryon number symmetry $Y_{B}\equiv \frac{1}{3V_s}\sum_{\mu=1,2,3}
\Bigg \langle (\sum_{x}B_{\mu}(x))^{2} \Bigg \rangle
\label{yb}$
where
$B_{\mu}(x)= \left[ b_\mu(x) \right]$.

In addition to these observables, the correlation lengths $M_{\pi}$ and $M_{B}$ associated with the chiral and diquark 
two point functions are also measured by calculating the temporal and 
spatial decays of the related two point functions respectively.

\section{Results}
\label{results}

\subsection{The Low Energy Effective Theory of the Model}
\label{EFT}

In order to study the finite temperature chiral singularities of the model, we have done extensive lattice simulations at a fixed finite temperature in the broken phase with a variety of different quark masses $m$. Further, since from our lattice calculations in the chiral limit, we found that the transition temperature $T_{C} \sim 2.9285$ \cite{FJ:2006ab}, hence we focus on the chiral singularities at $T=2.9$, a temperature which is below and close to $T_{C}$. 

\begin{figure}
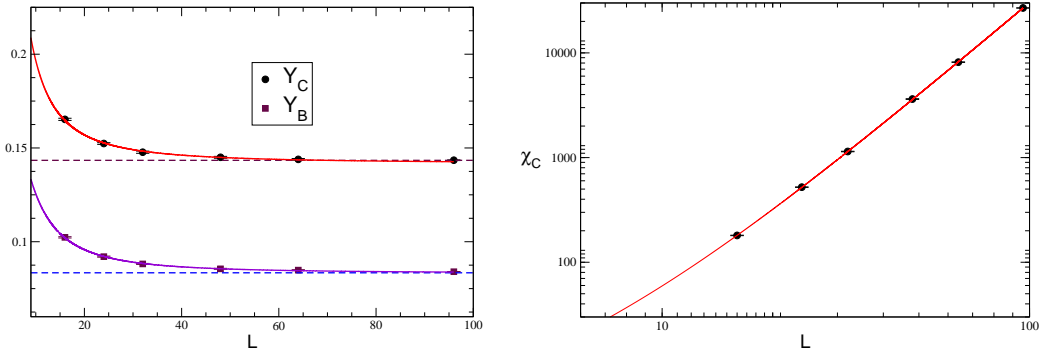

\hbox{
\includegraphics[width=0.4215\textwidth]{FB_t2.9.eps}
~
~
\includegraphics[width=0.4575\textwidth]{S_t2.9.eps}
}
\caption{\label{fig.1}(Left) This figure shows $Y_{C}$ and $Y_{B}$ as a function of box size $L$. The solid lines are the fits of $Y_{C}$ and $Y_{B}$ to formula (5.2). These fits shows that $F_{\pi}^2 = 0.1434(8)$, $F_{B}^2 = 0.1251(7)$. (Right) This figure shows $\chi_{C}$ as a function of box size $L$. The $\Sigma$ can be computed by fitting $\chi_{C}$ to formula (5.3) (solid line) and is given by $\Sigma = 0.4146(8)$.}
\end{figure}

Before describing the details of our lattice results, we review the low energy effective theory of the model. 
For small quark masses, it is possible to argue that the effective action $S_{h}$ of the model is given by:
\begin{equation}
S_{h}= \int d^{3}x \Bigg \{ \frac{F_{B}^{2}}{2}(\partial_{\mu}\vec{S})\cdot(\partial_{\mu}\vec{S})+\frac{F_{\pi}^{2}}{2}(\partial_{\mu}\vec{u})\cdot(\partial_{\mu}\vec{u})-\Sigma \vec{S}\otimes\vec{u}\cdot\vec{h} \Bigg \}.
\label{chpt1}
\end{equation}
where $\vec{S}$ is an unit 3-vector, $\vec{u}$ is an unit 2-vector and $\vec{h}$ is given by $\vec{h}=2m(1,0,0)\otimes(1,0)$. 
The $F_{\pi}^2$, $F_{B}^2$ and $\Sigma$ are the low energy constants of the effective theory \footnote{In the following, we will call $F_{\pi}$ the pion decay constant and $\Sigma$ the 
condensate}. Further, these low energy constants can be computated from our lattice
calculations in the massless theory. For example, $F_{\pi}^2$ and $F_{B}^2$ can be calculated by fitting $Y_{C}$ and $Y_{B}$ to their finite size scaling (FSS)
formulae respectively \cite{FJ:2006ab}:
\begin{equation}
\begin{array}{ll}
Y_{C} = F_{\pi}^{2} + a'L^{-1} + a''L^{-2}, \cr 
Y_{B}= \frac{2}{3} F_{B}^{2} + b'L^{-1} + b''L^{-2}.
\end{array}
\label{chpt2}
\end{equation}
where $a'$, $a''$, $b'$ and $b''$ are some constants. Further, using the extrapolated $F_{\pi}^2$ and $F_{B}^2$, $\Sigma$ can be obtained from the FSS formula of $\chi_{c}$:
\begin{equation}
\chi_{C}=L^{3}\frac{\Sigma^{2}}{6}\Bigg[1+\frac{\beta_{1}}{L}(\frac{1}{F_{\pi}^{2}}+\frac{2}{F_{B}^{2}})\Bigg]+a_{1}L.
\label{chpt3}
\end{equation}
where $a_{1}$ is some constant depending on the low energy constants of the theory. By this method, at $T = 2.9$, we found 
that $F_{\pi}^2 = 0.1434(8)$, $F_{B}^2 = 0.1251(7)$ and $\Sigma = 0.4146(8)$ (Fig.\ref{fig.1}). All the $\chi^2/DOF$ of fits in above and the following discussions are less than $1$. On the other hand, these low energy constants can also
be computed from our lattice simulations with massive quarks by performing the chiral extrapolation in the appropriate regime. In the
follwing, we will discuss this procedure of chiral extrapolations in detail.

\subsection{The Pion Decay Constant}

\begin{figure}
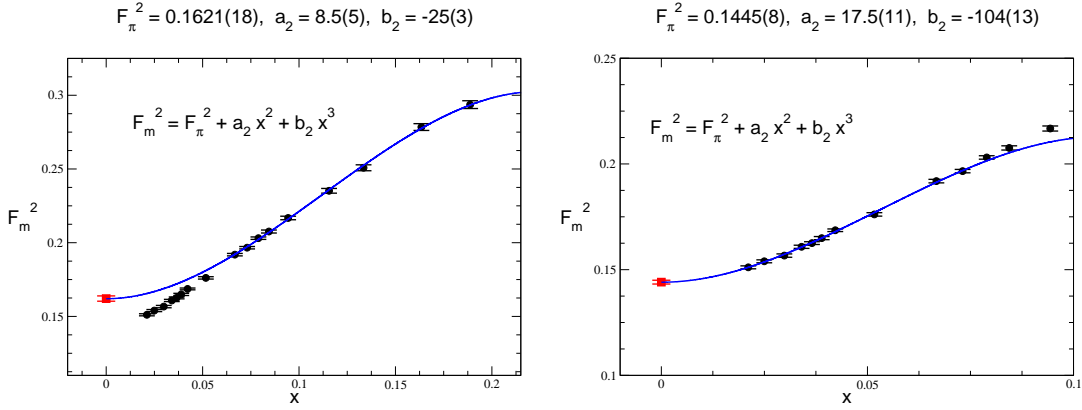

\hbox{
\includegraphics[width=0.454\textwidth]{3d_totalF.eps}
~
~
\includegraphics[width=0.46\textwidth]{3d_totalF1.eps}
}
\caption{(Left) This figure shows $F_{m}^2$ as a function of $x$. The solid line is the fit of $F_{m}^2$ with $0.06672 \le x \le 0.1887$ to the formula described in the text. This fit shows that $F_{\pi}^2 = 0.1621(18)$ which is not consistent with our $\epsilon$ regime lattice calculations. (Right) The fit of $F_{m}^2$ with $ 0.0211 \le x \le 0.0731$ shows $F_{\pi}^2 = 0.1445(8)$ which matchs nicely with the prediction. The most left data points (red squares) in both figures are the extrapolated results described in the text and are added to both figures by hands.}
\label{fig.2}
\end{figure}

In the regime where the condition $LM_{\pi} \gg 1$ is satisfied ($M_{\pi}$ is the pion mass), one would expect that the observables such as
the pion decay cosntant and condensate satisfy the expansion \cite{Hasenfratz:1989pk,BT1:2003ab}:
\begin{equation}
\langle O \rangle = a_{0} + b_{0}x + c_{0}x^2 + d_{0}x^3 + O(x^4).
\label{chpt4}
\end{equation}
where $x$ is a dimensionless parameter and is given by $x=\frac{\sqrt{2\Sigma m}}{4 \pi F_{\pi}^3}$, $a_{0}$ is the same observable obtained from the calculations with zero mass (see above) and $\sqrt{m}$ is the power-like singularity arising due to the infrared pion physics \cite{DJZ1:2003ab}. Further, from the effective lagrangian given by (\ref{chpt1}),
one can argue that the coefficient $b$ in (\ref{chpt4}) vanishes for the pion decay constant \cite{Hasenfratz:1989pk,BT1:2003ab}. Indeed, in the presence of massive quarks with mass $m$, if the pion decay constant $F_{m}^2$ is extrapolated by $F_{m}^2 = \lim_{L \to \infty} Y_{C}$, then we found that the $x$-dependent prediction $F_{\pi}^2 + c_{1}x^2 + d_{1}x^3$ of the pion decay constant is captured nicely by these extrapolated $F_{m}^2$. However, if we focus only on these
$F_{m}^2$ with $0.06672 \le x \le 0.1887$, then the fit shows that $F_{\pi}^2 = 0.1621(18)$ (left figure in Fig.\ref{fig.2}) which is obviously not consistent  with $F_{\pi}^2=0.1434(8)$ from our lattice calculations with massless 
quarks. On the other hand, if we restrict ourselves to the $F_{m}^2$ with $0.0211 \le x \le 0.0731$, then the same analysis shows that $F_{\pi}^2 = 0.1445(8)$ (right figure in Fig.\ref{fig.2}) which agrees nicely with its predicted value. If we fit $F_{m}^2$ to $F_{\pi}^2 + b_{1}x + c_{1}x^2 + d_{1}x^3$ with  $0.0211 \le x \le 0.0731$. Then the fit shows that  $F_{\pi}^2 = 0.1413(48)$ and $b_{1} = 0.24(35)$. Again both of them are consistent with the predictions $F_{\pi}^2=0.1434(8)$ and $b_{1} = 0$     

\subsection{The Condensate}

\begin{figure}
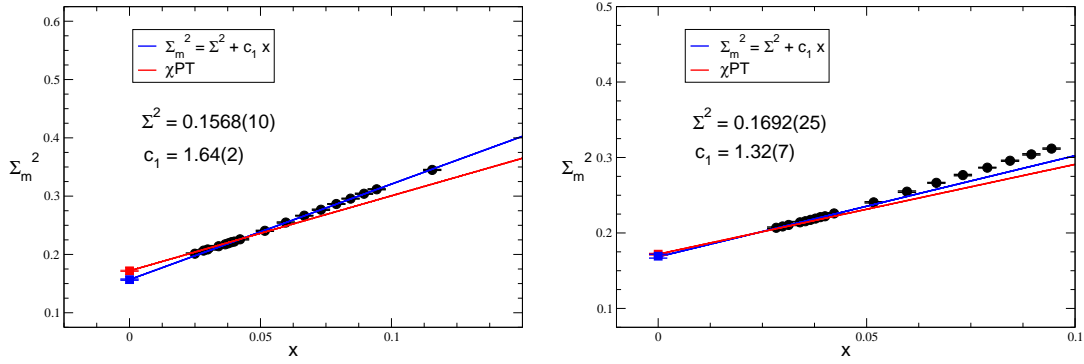

\hbox{
~
\includegraphics[width=0.454\textwidth]{3d_totalSigma.eps}
~
~
\includegraphics[width=0.46\textwidth]{3d_totalSigma1.eps}
}
\caption{(Left) This figure shows $\Sigma_{m}^2$ as a function of $x$. The blue solid line is the fit described in the text with $0.05168 \le x \le 0.1156$ and the red solid line is the prediction of $\chi$PT. The fit indicates $\Sigma^2 = a_{2} = 0.1568(10)$ and $b_{2} = 1.64(2)$. Neither of them match our predictions. (Right)  The fit of $\Sigma_{m}^2$ with $ 0.02497 \le x \le 0.05168$ shows $\Sigma^2 = 0.1692(25)$ and $b_{2} = 1.32(7)$ both of which match nicely with the predictions. Again, the blue solid line is the fit described in the text and the red solid line is the prediction of $\chi$PT. In both figures, the data points at $m=0$ are obtained from the $\epsilon$ regime calculations (red squares) and the extrapolation results (blue squares). They are added to the figures by hands.}
\label{fig.3}
\end{figure}

In terms of the dimensionless parameter $x$, the condensate $\Sigma_{m}$ obtained from our lattice simulations with massive quarks, is given by $\Sigma_{m}^2 = a_{2} + b_{2}x + O(x^2)$. Further, it can be shown that $a_{2} = \Sigma^2$ and
$b_{2} = -(\frac{\sqrt{2\Sigma}}{4 \pi F_{\pi}^3})^{-1}(\frac{2\Sigma^2}{F_{\pi}^2}G_{\pi}(0) + \frac{4\Sigma^2}{F_{B}^2}G_{B}(0))$, where $G_{i}(0) = \frac{-1}{4\pi}\sqrt{\frac{2\Sigma}{F_{i}^2}}$ \cite{BT1:2003ab}.
Using the $F_{\pi}^2$, $F_{B}^2$ and $\Sigma$ calculated in the massless theory, $a_{2}$ and $b_{2}$ are given by $a_{2} = 0.1719(10)$ and $b_{2} = 1.19(9)$ respectively. Again this prediction is 
captured nicely with the extrapolated $\Sigma_{m}^2$ which is obtained through $\Sigma_{m}^{2} = \lim_{L \to \infty}\chi_{C}/L^3$ if we include data at small $x$. However, as what has been already shown in the case of pion decay constant, the fitting results obtained from $\Sigma_{m}^2$ with larger $x$ dose not show a consistent result (left figure in Fig.\ref{fig.3}). Only with the $\Sigma_{m}^2$ from the smaller $x$ regime, the fit would show a satisfactory result, namely $a_{2}$ and $b_{2}$ calculated from the fit match nicely with our predictions (right figure in Fig.\ref{fig.3}).      

\section{Discussion and Conclusions}

We have used an efficient algorithm to study the chiral singularities 
of a two color lattice QCD model at a fixed finite temperature and have succeeded in understanding the low energy physics of the model in 
terms of a effective theory with a few low energy constants. In particular we found that the relation $\lim_{L \to \infty}\lim_{m \to 0}O = \lim_{m \to 0}\lim_{L \to \infty}O$ is satisfied, where $O$ is either the pion decay constant or the condensate. However in order to
see consistency, small quark masses are needed. In addition to the pion decay constant and the condensate, we also have studied the pion and diquark masses which will be
discussed elsewhere.

This work is done in collaboration with Shailesh Chandrasekharan and we thank Prof. T. Mehen, Prof. R. Springer, Prof. U.-J. Wiese and Dr. B. Tiburzi for helpful comments. This work was supported in part by the Department of Energy (DOE) grant DE-FG-02-05ER41368.

\end{document}